\documentclass{rmf-d}
\usepackage{nopageno,rmfbib,multicol,times,epsf,amsmath,amssymb,cite}
\usepackage[latin1]{inputenc}
\usepackage[]{caption2}
\usepackage{graphics}

\usepackage[demo]{graphicx}
\usepackage{lipsum}
\usepackage{wrapfig, blindtext}
\usepackage{comment}
\usepackage{tabularx}
\usepackage{float}

\def\rmfcornisa{Research OR Education of Physics \hfill\rmf\ {\bf ??} (*?*) ???--???
\hfill MES? A\~NO?}

\def\rmfcintilla{{\it Rev.\ Mex.\ Fis.\/} {\bf ??} (*?*) (????) ???--???}
\clearpage \rmfcaptionstyle 
\begin{document}
\title{Application of the weak-binding relation with range correction
\vspace{-6pt}}
\author{Tomona Kinugawa}
\address{Department of Physics, Tokyo Metropolitan University, Hachioji 192-0397, Japan}
\author{Tetsuo Hyodo}
\address{Department of Physics, Tokyo Metropolitan University, Hachioji 192-0397, Japan}
\author{ }
\address{ }
\author{ }
\address{ }
\author{ }
\address{ }
\author{ }
\address{ }
\maketitle
\recibido{day month year}{day month year
\vspace{-12pt}}
\begin{abstract}
\vspace{1em} 
The weak-binding relation is a useful tool to study the internal structure of hadrons from the observable quantities. We introduce the range correction in the weak-binding relation for the system having a sizable magnitude of the effective range, and show that the applicability of the weak-binding relation can be enlarged by the range correction. Thanks to the low-energy universality, the weak-binding relation can be used to study the structure of shallow bound states in any systems with different length scales. We apply the weak-binding relation to actual systems, including hadrons, hypernuclei, and atoms and show the importance of the range correction.
\vspace{1em}
\end{abstract}
\keys{Weak-binding relation; range correction; structure of hadrons \vspace{-4pt}}
\pacs{03.65.Ge; 03.65.Nk; 14.40.-n; \vspace{-4pt}}
\begin{multicols}{2}

%%%%%%%%%%%%%%%%%%%%%%%%%%%%%%%%%%%%%%%%%%%%%%%%%%%%%%%%%%%%%%%%%%%%%%%%
\section{Introduction}
%%%%%%%%%%%%%%%%%%%%%%%%%%%%%%%%%%%%%%%%%%%%%%%%%%%%%%%%%%%%%%%%%%%%%%%%

% exotic hadrons
It has been generally considered that most of the observed hadrons~\cite{ParticleDataGroup:2020ssz} consist of $\bar{q}q$ mesons or $qqq$ baryons, because of the absence of the quantum number exotic states~\cite{Hyodo:2020czb} and the phenomenological success of the constituent quark models for the description of the excited hadron spectra. Since the observation of $X(3872)$ by the Belle collaboration~\cite{Belle:2003nnu}, however, there have been growing evidence that the exceptions of $\bar{q}q/qqq$ classifications seem to exist. Such candidates are called exotic hadrons, and their internal structures are intensively studied theoretically and experimentally~\cite{Hosaka:2016pey,Guo:2017jvc,Brambilla:2019esw}.

% weak-binding relation
The investigation of the internal structure of hadrons has some similarity with the discussion in 1960s to distinguish the elementary particles from the composite ones. In that context, Weinberg studied the structure of the deuteron, showing that the deuteron is a composite system of the proton and neutron from the observables~\cite{Weinberg:1965zz}. This idea has been re-evaluated recently, and the weak-binding relation has been used to study the compositeness of hadrons~\cite{Kamiya:2015aea,Kamiya:2016oao}. The weak-binding relation is useful in hadron physics, particularly because the structure of hadrons can be determined from a few observables, without having the detailed knowledge of the interaction potential. 

% this work 
We have introduced the range correction in the weak-binding relation by considering the effective range parameter, and demonstrated that the applicability of the relation can be extended by the range correction~\cite{Kinugawa:2021ybb,Kinugawa:2021ykv}. Here we present the application of the weak-binding relation to the actual systems in which the range correction is important. Because the weak-binding relation is based on the low-energy universality~\cite{Braaten:2004rn,Naidon:2016dpf}, it is applicable to any system as long as the scattering length is sufficiently larger than other length scales in the system. We choose some examples of hadrons, nuclei, and atoms that satisfy this requirement, and study the structure of bound states using the weak-binding relation.

%%%%%%%%%%%%%%%%%%%%%%%%%%%%%%%%%%%%%%%%%%%%%%%%%%%%%%%%%%%%%%%%%%%%%%%%
\section{Weak-binding relation}
%%%%%%%%%%%%%%%%%%%%%%%%%%%%%%%%%%%%%%%%%%%%%%%%%%%%%%%%%%%%%%%%%%%%%%%%

% weak-binding relation
Consider the two-body system with the scattering length $a_{0}$ having a shallow bound state with the binding energy $B$. The compositeness $X$ is defined as the probability of finding the two-body scattering states in the wavefunction of the bound state. These quantities follow the weak-binding relation~\cite{Kamiya:2015aea,Kamiya:2016oao}
\begin{align}
\label{eq:wbr}
a_0&=R\left\{\frac{2X}{1+X}+\mathcal{O}\left(\frac{R_{\rm typ}}{R}\right)\right\},
\end{align}
where $R=1/ \sqrt{2\mu B}$ is the length scale by the binding energy with the reduced mass of the system $\mu$. $R_{\rm typ}$ is the typical length scale of the two-body system, which will be discussed in detail below. The weak-binding relation~\eqref{eq:wbr} shows that the compositeness $X$ can be determined from the observables $a_{0}$ and $R$ (or $B$) when the binding energy is so small that the correction terms of $\mathcal{O}\left(R_{\rm typ}/R\right)$ are negligible. 

% $R_{\rm int}$
Let us now discuss the length scale $R_{\rm typ}$. In Refs.~\cite{Kamiya:2015aea,Kamiya:2016oao}, the relation~\eqref{eq:wbr} is derived using the nonrelativistic  effective field theory with contact interactions, and the correction terms appear from the finite cutoff $\Lambda$ of the momentum integration. Because the inverse of the momentum cutoff corresponds to the length scale where the contact interaction theory is applicable, it gives the interaction range of the microscopic theory, $R_{\rm int}\sim 1/\Lambda$. From this observation, the length scale $R_{\rm typ}$ was estimated in Refs.~\cite{Kamiya:2015aea,Kamiya:2016oao} by the interaction range as,
\begin{align}
\label{eq:correction-old}
R_{\rm typ}=R_{\rm int}. 
\end{align}

%%%%%%%%%%%%%%%%%%%%%%%%%%%%%%%%%%%%%%%%%%%%%%%%%%%%%%%%%%%%%%%%%%%%%%%%
\section{Range correction}
%%%%%%%%%%%%%%%%%%%%%%%%%%%%%%%%%%%%%%%%%%%%%%%%%%%%%%%%%%%%%%%%%%%%%%%%

% $R_{\rm eff}$
We notice that the interaction range $R_{\rm int}$ is not the only length scale of the system. For instance, the effective range $r_{e}$ introduces the additional length scale in the scattering amplitude. Let us define the length scale $R_{\rm eff}$ as the largest length scale in the effective range expansion of the scattering amplitude. In usual cases, $R_{\rm eff}$ is of the same order with $R_{\rm int}$. This is however not always guaranteed, because $R_{\rm eff}$ is essentially different from $R_{\rm int}$. The cutoff $\Lambda$ is introduced for the integration of the off-shell momentum. In other words, $R_{\rm int}$ expresses the length scale in the off-shell nature of the system. On the other hand, $R_{\rm eff}$ is defined through the effective range expansion of the on-shell scattering amplitude, and hence expresses the length scale of the on-shell nature of the system.

% range correction
According to the discussion in Refs.~\cite{Kinugawa:2021ybb,Kinugawa:2021ykv}, we introduce the range correction in the weak-binding relation by the redefinition of the length scale $R_{\rm typ}$ in the correction term as
\begin{align}
\label{eq:correction-new}
R_{\rm typ}=\max\left\{R_{\rm int},R_{\rm eff}\right\},
\end{align}
namely, the larger one among $R_{\rm int}$ and $R_{\rm eff}$. In fact, using the effective range model~\cite{Braaten:2007nq}, we explicitly construct a system which does not follow the weak-binding relation with the correction term~\eqref{eq:correction-old}, but is consistent only with the improved correction term~\eqref{eq:correction-new}~\cite{Kinugawa:2021ybb,Kinugawa:2021ykv}. We also show that the range correction should be introduced in the correction term, by analyzing the origin of the effective range~\cite{Kinugawa:2021ykv}.

% applicability
To examine the applicability of the weak-binding relations, we perform numerical calculations with the effective range model with a finite cutoff~\cite{Braaten:2007nq}. In this model, the exact value of the compositeness of the bound state is fixed as $X=1$ because there is no channel coupling to other states. In addition, the finite cutoff $\Lambda$ provides the scale of the interaction range $R_{\rm int}$, and the values of the scattering length $a_{0}$ and the effective range $r_{e}$ can be arbitrarily chosen by adjusting the bare parameters. Hence, we can examine the applicability of the weak-binding relation by varying the $R_{\rm int}$ and $R_{\rm eff}=|r_{e}|$ in the model. We find that the improved relation with the correction term~\eqref{eq:correction-new} has larger applicability than that with the previous relation with Eq.~\eqref{eq:correction-old}~\cite{Kinugawa:2021ybb}.

%%%%%%%%%%%%%%%%%%%%%%%%%%%%%%%%%%%%%%%%%%%%%%%%%%%%%%%%%%%%%%%%%%%%%%%%
\section{Application}
%%%%%%%%%%%%%%%%%%%%%%%%%%%%%%%%%%%%%%%%%%%%%%%%%%%%%%%%%%%%%%%%%%%%%%%%

%------------------------------
\subsection{Bound states and two-body systems}

% systems
We now apply the weak-binding relations to the bound states summarized in Table~I. As a representative example of experimentally established states, we consider the deuteron ($d$) near the threshold of the proton-neutron ($pn$) system. Among recently observed exotic hadron candidates, we choose $X(3872)$ near the $D^{0}\bar{D}^{*0}$ threshold (in this notation, linear combination with the charge conjugate pair is implicit). In recent lattice QCD studies, existence of a shallow bound state is predicted in several systems. Here we study the structure of the $N\Omega$ dibaryon~\cite{HALQCD:2018qyu} and of the $\Omega\Omega$ dibaryon~\cite{Gongyo:2017fjb}. Weakly bound states can also be found in nuclear systems. Let us choose the hypertriton ${}^{3}_{\Lambda}{\rm H}$ near the $d\Lambda$ threshold. We further consider the atomic systems. A well known example of shallow bound states is the ${}^{4}{\rm He}$ dimer. Masses of hadrons are taken from PDG~\cite{ParticleDataGroup:2020ssz} except for the $N\Omega$ and $\Omega \Omega$ dibaryons where the lattice hadron masses are used, and the mass of 
the ${}^{4}{\rm He}$ atom is taken from Ref.~\cite{Wang:2021xhn} for the calculation of the reduced mass of the system.

% Coulomb interaction 
Comments on the Coulomb interaction and the decay effects are in order. Most of the two-body systems considered are free from the Coulomb interaction, because they include at least one electrically neutral particle. For the $N\Omega$ system, we can choose the neutron $n$ to avoid the Coulomb effect. While the Coulomb interaction is unavoidable in the $\Omega\Omega$ system, here we use the values of the scattering parameters obtained only by the strong interaction, and study the property of the dibaryon in the absence of the Coulomb interaction. Note also that the $\Omega\Omega$ dibaryon is shown to remain bound even if the Coulomb repulsion is included~\cite{Gongyo:2017fjb}. 

% decay effect
While the weak-binding relation~\eqref{eq:wbr} should be applied to stable bound states, $X(3872)$ and the $N\Omega$ dibaryon are the unstable states which decays via the strong interaction to the lower energy coupled channels. The experimental decay width of $X(3872)$ is experimentally determined as $1.19\pm 0.21\ $MeV~\cite{ParticleDataGroup:2020ssz} which is significantly smaller than the typical decay width of the excited hadrons ($ \mathcal{O}(100)$ MeV). The $N\Omega$ dibaryon has $J^{P}=2^{+}$ and decays into the $\Lambda \Xi$ and $\Sigma \Xi$ channels in $d$ wave. In fact, the decay width is obtained as $\sim 1$ MeV by the explicit calculation in the meson-exchange model~\cite{Sekihara:2018tsb}. We thus ignore the decay effects of the $X(3872)$ and the $N\Omega$ dibaryon, and apply the weak-binding relation for the stable bound states~\eqref{eq:wbr}.

% $a_{0}, r_{e}, B$
Next, we summarize the scattering length $a_{0}$ and the effective range $r_{e}$ of the two-body system, and the binding energy of the bound state $B$ shown in Table~I. For the deuteron, $a_{0}$ and $r_{e}$ of the $pn({}^{3}{\rm S}_{1})$ scattering, together with the deuteron binding energy $B$ are taken from the CD-Bonn potential~\cite{Machleidt:2000ge}. The $D^{0}\bar{D}^{*0}$ scattering parameters and the binding energy of $X(3872)$ are taken from Ref.~\cite{Esposito:2021vhu} based on the Flatt\'e analysis by the LHCb collaboration~\cite{LHCb:2020xds} (note, however, the ambiguity of the determination of $r_{e}$ discussed in Ref.~\cite{Baru:2021ldu}). The parameters $a_{0},r_{e}$ and $B$ for the $N\Omega$ and $\Omega\Omega$ dibaryons are taken from the lattice QCD analysis~\cite{HALQCD:2018qyu,Gongyo:2017fjb}. In both cases, $a_{0},r_{e}$ and $B$ are determined by solving the Schr\"odinger equation with the lattice hadron masses which are slightly heavier than the physical ones ($m_{N}^{\rm lat}=955$ MeV and $m_{\Omega}^{\rm lat}=1712$ MeV). For the $\Omega\Omega$ dibaryon, we use the binding energy $B$ obtained without Coulomb potential, as discussed above. The hypertriton binding energy from the $\Lambda d$ threshold (separation energy into $\Lambda d$) is taken from the emulsion data~\cite{Juric:1973zq}. We adopt the scattering parameters $a_{0},r_{e}$ from the effective field theory analysis~\cite{Hammer:2001ng}. As for the ${}^{4}{\rm He}$ dimer, we use $a_{0},r_{e},B$ shown in Ref.~\cite{Kievsky:2012ss} which are obtained by solving the LM2M2 potential.

% $R_{\rm int}$
Finally, we estimate the interaction length scale $R_{\rm int}$ in each system. In contrast to $a_{0},r_{e}$ and $B$, the interaction range is not a direct observable. It is therefore needed to estimate $R_{\rm int}$ from the longest possible length scale by considering the underlying interaction mechanisms. For the deuteron $(d)$ and 

\end{multicols}
\begin{table}[htb]
\label{tab:applicable-systems}
\centering
  \caption{Properties of the bound states near the threshold of the two-body system. The scattering length $a_{0}$, the effective range $r_{e}$, and the interaction range $R_{\rm int}$ of the two-body system are shown. The binding energy of the bound state $B$ is measured from the threshold of the two-body system in this table. B.R. stands for the Bohr radius.} 
  \vspace*{0.3cm}
\renewcommand{\tabcolsep}{1.35pc}
  \begin{tabular}{ccccccc}  
  \hline
    Bound state & Two-body system & $a_0$ & $r_e$ & $R_{\rm int}$ & $B$ \\ \hline 
    $d$ & $pn({}^{3}{\rm S}_{1})$ & 5.42 fm & 1.75 fm & $1.43$\ fm & $2.22$ MeV \\ 
    $X(3872)$ & $D^{0}\bar{D}^{*0}$ & 28.5 fm & $-5.34$ fm & $1.43$\ fm & 18 keV \\ 
        $N\Omega $ dibaryon & $n\Omega ({}^{5}{\rm S}_{2})$ & 5.30\ fm &1.26\ fm & $0.676$ fm & $1.54$\ MeV  \\ 
        $\Omega \Omega$ dibaryon & $\Omega \Omega({}^{1}{\rm S}_{0})$ & 4.6\ fm&1.27\ fm & $0.949$ fm & $1.6$\ MeV \\  
        ${}^{3}_{\Lambda}{\rm H}$ & $d\Lambda$ & 16.8\ fm &2.3\ fm & $4.31$ fm & $0.13$\ MeV \\
         ${}^{4}{\rm He}$ dimer & ${}^{4}{\rm He}{}^{4}{\rm He}$ &  189\ B.R.&13.8\ B.R. & $10.2$ B.R. & $1.30$\ mK \\   
  \hline
  \end{tabular}
\end{table}
\begin{multicols}{2}

\noindent
$X(3872)$, the pion exchange is possible in the corresponding two-body systems. Because the pion is the lightest hadron which can be exchanged, we estimate the interaction range $R_{\rm int}$ by the pion Compton wavelength $R_{\rm int}\sim 1/m_{\pi}$. In Ref.~\cite{HALQCD:2018qyu}, the $N\Omega$ potential is parametrized by Gaussian $+({\rm Yukawa})^{2}$ form with the lattice pion mass $m_{\pi}^{\rm lat}=146$ MeV for the Yukawa term. Because this potential is exponentially suppressed beyond the distance $1/2m_{\pi}^{\rm lat}$, we estimate $R_{\rm int}\sim 1/2m_{\pi}^{\rm lat}=0.676$ fm (the range of the Gaussian part is $\sim 0.1$ fm). As for the $d\Lambda$ system, we estimate the interaction range $R_{\rm int}$ by the radius of the deuteron. The interaction range of the ${}^{4}{\rm He}$ dimer is estimated by the van der Waals length $R_{\rm int}\sim l_{\rm vdW}=(mC_{6}/\hbar^{2})^{1/4}$ with the coefficients $C_{6}$ calculated in Ref.~\cite{Yan}.

%------------------------------
\subsection{Evaluation of the compositeness}

% length scale comparison
In Table~I, we find that the scattering length is the largest length scale, $a_{0}>|r_{e}|,R_{\rm int}$ in each system. This justifies the use of the weak-binding relation to study the compositeness of these bound states, even though the length scale of the strong interaction (fm) is completely different from that in the atomic system (\AA). In addition, except for the hypertriton, the magnitude of the effective range is larger than the estimated interaction range $R_{\rm int}$. 

% applicability
Comparing with the applicability of the weak-binding relations with respect to $R_{\rm eff}$ and $R_{\rm int}$, we find that $X(3872)$ lies in the region where only the improved weak-binding relation with Eq.~\eqref{eq:correction-new} can be applied. Also, the $N\Omega$ dibaryon lies close to the boundary of the applicability of the previous relation with Eq.~\eqref{eq:correction-old}. This implies that the range correction is particularly important for these cases, and the weak-binding relation with the previous correction term~\eqref{eq:correction-old} might fail to estimate the compositeness.

% uncertainty result
Following the uncertainty estimation procedure proposed in Ref.~\cite{Kamiya:2016oao}, we apply the weak-binding relations to the bound states in Table~I. In Table~II, we show the estimated values of the compositeness of the bound state $X$ including the uncertainties with the typical length scale $R_{\rm typ}=R_{\rm eff}=|r_{e}|$ and $R_{\rm typ}=R_{\rm int}$, separately. The previous weak-binding relation with Eq.~\eqref{eq:correction-old} gives the results in the column $R_{\rm typ}=R_{\rm int}$, while for the improved one with Eq.~\eqref{eq:correction-new}, the results with $R_{\rm typ}=R_{\rm eff}$ are adopted except for ${}^{3}_{\Lambda}{\rm H}$. We find that the lower bound of the compositeness of the $N\Omega$ dibaryon is 1.04 for $R_{\rm typ}=R_{\rm int}$. This contradicts with the definition of the compositeness of the bound state $0\leq X\leq 1$. Namely, the previous weak-binding relation ($R_{\rm typ}=R_{\rm int}$) cannot be used for the $N\Omega$ dibaryon. On the other hand, the lower bound is 0.801 with $R_{\rm typ}=R_{\rm int}$, and therefore the improved relation with $R_{\rm typ}=R_{\rm eff}$ provides the compositeness $X$ consistent with the definition. 

% final result
Taking into account the definition of the compositeness $0\leq X \leq 1$, the results of the weak-binding relation with the improved correction term~\eqref{eq:correction-new} are shown in the fourth column of Table~II. The results show that the composite component dominates the internal structure of the bound states studied here. In particular, more than 90\% of the ${}^{4}{\rm He}$ dimer consists of the composite component. On the other hand, while the composite dominance holds for $X(3872)$, the lower bound around 0.5 indicates the nonnegligible mixing of the components other than the $D^{0}\bar{D}^{*0}$ one. 

\tabletopline\vspace{2pt}\lilahf{\sc Table II.\ {\rm Estimation of the compositeness of bound states $X$ by the weak-binding relation with $R_{\rm typ}=R_{\rm eff}$ (second column) and $R_{\rm typ}=R_{\rm int}$ (third column). The fourth column shows the results by the improved correction term~\eqref{eq:correction-new} together with the constraint from the definition $0\leq X \leq 1$.}}
\begin{center}
\small{\renewcommand{\arraystretch}{1.3}
\begin{tabular}{cccc}  \hline
    Bound state & $R_{\rm typ}=R_{\rm eff}$ & $R_{\rm typ}=R_{\rm int}$ & This work \\ \hline 
    $d$ & $1.68^{+3.19}_{-0.943}$ & $1.68^{+2.14}_{-0.823}$ & $0.738\leq X\leq 1$ \\ 
    $X(3872)$ & $0.743^{+0.282}_{-0.213}$ & $0.743^{+0.0675}_{-0.0627}$ & $0.530\leq X\leq 1$ \\ 
    $N\Omega$ dibaryon & $1.40^{+1.20}_{-0.600}$ & $1.40^{+0.523}_{-0.364}$ & $0.801\leq X\leq 1$ \\ 
    $\Omega \Omega$ dibaryon & $1.56^{+1.95}_{-0.773}$ & $1.56^{+1.22}_{-0.626}$ & $0.791\leq X\leq 1$ \\
    ${}^{3}_{\Lambda}{\rm H}$ & $1.35^{+0.531}_{-0.366}$ & $1.35^{+1.241}_{-0.603}$ & $0.745\leq X\leq 1$ \\ 
    ${}^{4}{\rm He}$ dimer & $1.08^{+0.179}_{-0.152}$ & $1.08^{+0.129}_{-0.115}$ & $0.926\leq X\leq 1$ \\
     \hline
  \end{tabular}
}
\end{center}

%%%%%%%%%%%%%%%%%%%%%%%%%%%%%%%%%%%%%%%%%%%%%%%%%%%%%%%%%%%%%%%%%%%%%%%%
\section{Summary}
%%%%%%%%%%%%%%%%%%%%%%%%%%%%%%%%%%%%%%%%%%%%%%%%%%%%%%%%%%%%%%%%%%%%%%%%

Based on the discussion of the length scales, we propose the range correction in the weak-binding relation by modifying the correction term. From the hadrons, nuclei, and atomic systems, we list the weakly bound states with the sizable effective range. The compositeness of these bound states is quantitatively evaluated by using the weak-binding relations. It is shown that the range correction of the weak-binding relation is necessary for the meaningful estimation of the compositeness of the $N\Omega$ dibaryon. These resutls indicate the importance of the range correction for the study of the internal structure of hadrons.

\end{multicols}
\medline
\begin{multicols}{2}

\end{multicols}
\end{document}